\documentclass[preprint]{aastex}
\usepackage{epsfig}

\def\hypgeo{\mathop{_2{\rm F}_1}}

\newcommand{\GRB}{GRB~000301C}
\newcommand{\FL}{F_{\rm L}}
\newcommand{\FH}{F_{\rm H}}
\newcommand{\LL}{L_{\rm L}}
\newcommand{\LH}{L_{\rm H}}
\newcommand{\fH}{f_{\rm H}}
\newcommand{\fL}{f_{\rm L}}
\newcommand{\Ft}{{F_{\rm th}}}

\begin{document}
\title{Microlensing of Gamma Ray Bursts by Stars and MACHOs}
\author{Edward A. Baltz\altaffilmark{1} and Lam Hui\altaffilmark{2}}

\altaffiltext{1}{KIPAC, Stanford University, P.O.~Box~90450, MS 29, Stanford,
  CA 94309, {\tt eabaltz@slac.stanford.edu}}

\altaffiltext{2}{NASA/Fermilab Astrophysics Center, Fermi National Accelerator
Laboratory, PO Box 500, Batavia, IL 60510 {\tt lhui@fnal.gov}}

\begin{abstract}
The microlensing interpretation of the optical afterglow of \GRB\ seems naively
surprising, since a simple estimate of the stellar microlensing rate gives less
than one in four hundred for a flat $\Omega_\Lambda=0.7$ cosmology, whereas one
event was seen in about thirty afterglows.  Considering baryonic MACHOs making
up half of the baryons in the universe, the microlensing probability per burst
can be roughly $5\%$ for a GRB at redshift $z=2$.  We explore two effects that
may enhance the probability of observing microlensed gamma-ray burst
afterglows: binary lenses and {\it double} magnification bias.  We find that
the consideration of binary lenses can increase the rate only at the $\sim 15
\%$ level.  On the other hand, because gamma-ray bursts for which afterglow
observations exist are typically selected based on fluxes at widely separated
wavebands which are not necessarily well correlated (e.g.\ localization in
X-ray, afterglow in optical/infrared), magnification bias can operate at an
enhanced level compared to the usual single--bias case.  Using a simple model
for the selection process in two bands, we compute the enhancement to
microlensing rate due to magnification bias in two cases: perfect correlation
and complete independence of the flux in the two bands.  We find that existing
estimates of the slope of the luminosity function of gamma-ray bursts, while as
yet quite uncertain, point to enhancement factors of more than three above the
simple estimates of the microlensing rate.  We find that the probability to
observe at least one microlensing event in the sample of 27 measured afterglows
can be $3-4\%$ for stellar lenses, or as much as $25\,\Omega_{\rm lens}$ for
baryonic MACHOs.  We note that the probability to observe at least one event
over the available sample of afterglows is significant only if a large fraction
of the baryons in the universe are condensed in stellar--mass objects.
\end{abstract}

\keywords{gamma rays:bursts --- gravitational lensing -- binaries: general}

\section{Introduction}
The optical afterglow of \GRB, at a redshift of $z=2.034$, exhibited
variability prior to the break in the power law decline in its lightcurve.
This variability has been interpreted as gravitational microlensing (Garnavich,
Loeb \& Stanek 2000), as first proposed by Loeb \& Perna (1998).  While there
is no obvious intervening galaxy which might provide lenses in the form of
stars or MACHOs, we will take this possibility seriously.  Koopmans and
Wambsganss (2001) considered lensing by MACHOs (see also Wyithe \& Turner
(2002b)), but as we will explain in \S\ref{discussion} we find a rate of MACHO
lensing five times larger than theirs.  The rate of microlensing by stars is
much lower, and is naively of order one in three hundred GRBs at redshift two.
We note that microlensing is not the only possible explanation of this
lightcurve (Panaitescu 2001).  Several authors have considered using
microlensing to study the properties of the underlying afterglow (Granot \&
Loeb 2001; Gaudi \& Loeb 2001; Gaudi, Granot \& Loeb 2001; Ioka \& Nakamura
2001).  In this work we will only make statements about afterglows in the
aggregate.

Assuming that there is a fraction $f\sim0.1-0.2$ of dark matter in MACHOs, it
is not surprising that a lensing event was seen.  However, if the only
cosmological microlenses are stars, the observation of an event is somewhat
problematic.  Roughly thirty afterglows have been well--characterized (starting
with Metzger et al. (1997) and Kulkarni et al. (1998) and summarized in Frail
et al.\ (2001) and Bloom, Frail \& Kulkarni (2003)).  We investigate whether
magnification bias can make a significant difference, considering the two
extreme possibilities that the optical and gamma ray fluxes are perfectly
correlated or perfectly uncorrelated.  In Appendix~\ref{appdx:binary} we
consider binary lenses, as discussed by Mao \& Loeb (2001).  Neither of these
two effects seems to be able to increase the probability by a factor of ten.
We thus conclude that the observation of a second microlensing event by a star
in a sample twice as large is quite unlikely, though with a large MACHO
fraction the probability can be significant.

In \S\ref{physics} we summarize the basic physics involved in gamma ray burst
fireballs, and give an overview of the multiwavelength observations.  In
\S\ref{microlensing} we derive the cosmological microlensing rate (relegating
the details for binary lenses to Appendix~\ref{appdx:binary}).  In the
remaining sections (\S\ref{bias} -- \S\ref{sample}) we discuss the enhancement
of the lensing probability due to magnification bias (Turner, Gott \& Ostriker
1984).  Magnification bias is the effect that a given {\em detected} source is
more likely to be lensed than a source chosen randomly, simply because
magnified sources are easier to detect.  We will consider both the usual single
magnification bias, as well as double magnification bias which might be
relevant here because of the way gamma-ray burst afterglows are selected.  The
selection is complicated, involving gamma ray, X-ray and optical wavelengths.
We can only hope to capture the basic features.  In \S\ref{bias} \&
\S\ref{doublebias} we consider how the magnification bias affects bursts at a
single redshift, while in \S\ref{sample} we consider the magnification bias on
the sample of well--measured afterglows in Bloom et al. (2003).

We primarily consider the gravitational lensing caused by point mass lenses.
We define the distance to the source $D_s$, lens $D_\ell$, and
deflector--source distance $D_{\ell s}$.  These distances are all angular
diameter distances in physical units, and we take a flat cosmology where
$\Omega_m+\Omega_\Lambda=1$.  We define the Einstein radius and the Einstein
angle for the system as
\begin{equation}
\label{eq:R_E}
R_E=\sqrt{\frac{4GM}{c^2}\frac{D_\ell D_{\ell s}}{D_s}},\hspace{0.2in}
\theta_E=\frac{R_E}{D_\ell},
\end{equation}
where $M$ is the mass of the point lens.

\section{Physics and Observations of GRB Afterglows}
\label{physics}

The basic picture of a gamma-ray burst (GRB) afterglow is a fireball with a
relativistic jet with some small opening angle (e.g.\ Waxman 1997a).  The
observed afterglow light is emitted in a ring of opening angle $1/\gamma$,
where $\gamma$ is the Lorentz factor of the ejecta moving at velocity $v$.
This ring expands at an apparent rate $v_{\rm app}=\gamma v$, the so--called
superluminal expansion.  This superluminal expansion is what allows for rapid
microlens variability, even at cosmological distances (Loeb \& Perna 1998).

We take the radius of the afterglow ring in units of the Einstein radius from
Waxman (1997b) and Garnavich et al. (2000),
\begin{equation}
R(t)=0.50\left(\frac{E_{53}}{n_1}\right)^{1/8}(1+z_s)^{-5/8} \left(\frac{M_{\rm
lens}}{M_\odot}\right)^{-1/2} \left(\frac{D_\ell D_{\ell
s}/D_s}{c/H_0}\right)^{-1/2}\,t_{\rm day}^{5/8},
\end{equation}
where $E_{53}$ is the isotropic burst energy in units of $10^{53}$ ergs, and
$n_1$ is the ambient gas density in units of 1 cm$^{-3}$.  We will assume for
this work that we want a microlensing event to occur at some minimum time after
the GRB, to allow for optical observations.  To enforce this assumption we take
$R>0.2$, typically achieved roughly one tenth day after the burst at a redshift
of $z=2$.  We define the fractional apparent width of the emitting region $w$,
where the emission comes in the annulus between $R(1-w)$ and $R$.  In fact,
there is some emission throughout the interior of the ring, but the inner part
is less bright (Granot, Piran \& Sari 1999).

For our later calculation, we need to know the intrinsic (not apparent)
luminosity function of GRBs, at different wavebands.  This is a subject of much
uncertainty at the moment. Schaefer, Deng \& Band (2001) found that the
luminosity function $N(L)\equiv dn/dL$ at gamma-ray energies is a power law of
slope $-1.7$ at the faint end. This result is obtained by assuming a
lag--variability relation. Note that bursts of all durations are included in
their analysis, whereas bursts for which afterglows have been observed fall
into the long duration category (as localization of the bursts is typically
done with BeppoSAX observations which are insensitive to bursts with durations
less than about a second).  Frail et al. (2001) demonstrated using about 20
afterglows that the total energy output is remarkably constant from burst to
burst, about $5 \times 10^{50}$ ergs, most of which comes out in gamma-ray.
Bloom et al. (2003) confirmed this result with a larger sample of about 30
afterglows.  However, what is important to us is not the true total energy, but
rather the isotropic equivalent energy. It is the latter that determines the
effective luminosity of the burst (i.e. the observed fluence is proportional to
the isotropic equivalent energy divided by the luminosity distance
squared). The two are related via a geometrical factor determined by the
opening angle of the burst. The luminosity function is therefore determined by
the distribution of opening angles.  Frail et al. (2001) derived this
distribution using the observed jet break times. However, it is important to
keep in mind that the distribution of opening angles is obtained from a
flux-limited sample.\footnote{The sample is heterogeneous in nature, assembled
from observations by a variety of groups under different conditions. To the
extent that faint GRBs are less likely to be observed, one can regard the
sample as crudely flux-limited.} To relate that to the intrinsic distribution
of opening angles, and therefore luminosities, would require detailed modeling
of the selection, which is beyond the scope of this paper.\footnote{It is worth
noting that the luminosity function obtained directly from a flux limited
sample with no corrections is always biased against faint sources (even if the
source redshifts are known), typically driving the inferred function flatter
than the true one.}  Krumholz, Thorsett \& Harrison (1998) derived constraints
on the GRB luminosity function by matching the number-peak-flux relation in the
BATSE catalog (see also Rutledge, Hui \& Lewin 1995), as well as making use of
three bursts with known redshifts at the time. Assuming that the redshift
distribution follows the global star formation history (see Madau et al. 1996;
Madau, Pozzetti \& Dickinson 1998; Wijers et al. 1998), they find that a
power-law slope of $-1.6$ to $-2.05$ is consistent with observations (see
Fig.~1 in Krumholz et al. 1998).

The luminosity function at optical wavebands is even more uncertain, because of
the limited number of afterglows.  On theoretical grounds, one expects the
gamma-ray emission to be well correlated with X-ray, but not with the optical.
This is because the X-ray frequencies are generally above the various
characteristic frequencies of the relativistic fireball, such as e.g.\ the
synchrotron self--absorption frequency, whereas the (unknown) relative ordering
of the characteristic frequencies affects the optical flux significantly.  Some
observational evidence of this can be found in Freedman \& Waxman (2001) (see
also useful discussions in Kumar (2000) and Sari, Piran \& Narayan (1998)).
For our later investigation, we will examine both the case where the gamma-ray
/ X-ray and optical emission are well-correlated, and the case where they are
not. As far as the slope of the optical luminosity function is concerned, we
will assume, for simplicity and the lack of evidence to the contrary, it is the
same as that for the gamma-ray emission.

Observation of an optical afterglow has a complicated selection process, which
we will outline here.  First, the burst is detected in gamma-rays ($>20$ keV or
so) by BATSE, but it is at lower X-ray energies that the bursts are localized
by e.g.\ BeppoSAX.  After a localization, an attempt to measure the optical
afterglow can be made.  Thus, to detect an optical afterglow, there are
effectively three thresholds that must be exceeded.  The gamma-ray and X-ray
fluxes are believed to be well--correlated, but the correlation with the
optical flux is uncertain (Granot \& Sari 2002).

\section{Microlensing Rate}
\label{microlensing}

\begin{figure}
\epsfig{width=\textwidth,file=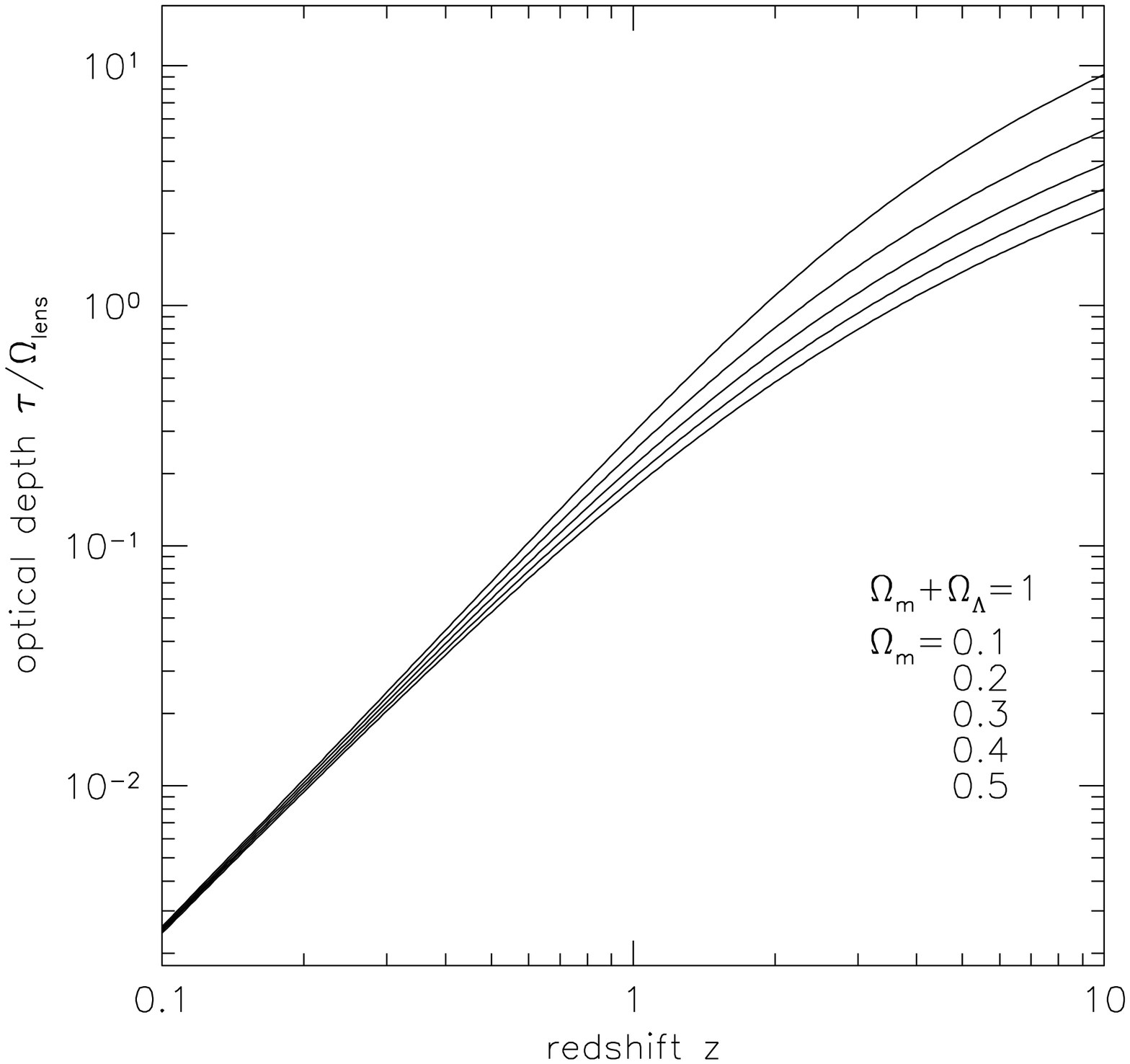}
\caption{Optical depth for microlensing for several values of $\Omega_m$ in a
flat cosmology.  We have divided out the $\Omega_{\rm lens}$ dependence.  The
curves from top to bottom correspond to $\Omega_m =$ 0.1 to 0.5 in steps of
0.1.  For our fiducial case of $\Omega_m=0.3$ and $z=2$, we find
$\tau=0.65\,\Omega_{\rm lens}$.}
\label{fig:tau}
\end{figure}

The microlensing optical depth is defined as the number of lenses within some
minimum impact parameter of the line of sight to an object.  The basic optical
depth, where the impact parameter is taken to be an Einstein radius, is given
by the following for a flat cosmology (Press \& Gunn 1973):
\begin{equation}
\tau_0=\frac{3}{2}{\Omega_{\rm lens}}\,\int_0^{z_s} 
\frac{dz_\ell\,(1+z_\ell)}{\sqrt{\Omega_m(1+z_\ell)^3+\Omega_\Lambda}}
\frac{\chi_\ell(\chi_s-\chi_\ell)}{\chi_s},
\end{equation}
where $\chi$ is the dimensionless coordinate distance, i.e.\ the luminosity
distance is $D_L=c\chi(1+z)/H_0$.  In Fig.~\ref{fig:tau} we plot the optical
depth as a function of redshift for several values of $\Omega_m$ in a flat
cosmology, dividing out the $\Omega_{\rm lens}$ dependence.  The quantity
$\Omega_{\rm lens}$ is the fraction of critical density today that is in
objects capable of causing microlensing.  We have assumed that the comoving
number density of such objects is constant -- if it decreases towards higher
redshifts, the optical depth will be lower.  For $\Omega_m=0.3$ and $z=2$, the
optical depth is $\tau_0=0.65\,\Omega_{\rm lens}$.  Fukugita, Hogan \& Peebles
(1998) estimated $\Omega_{\rm lens}$ in stars to be $\Omega_{\rm lens}\approx
0.0035$.  With this value, constant out to redshift $z=2$, the optical depth is
1/440 for $\Omega_m=0.3$, and is as large as 1/260 for $\Omega_m=0.1$, though
the latter value for the matter density is strongly disfavored.  For MACHOs, we
will simply assume that $\Omega_{\rm lens}=f\Omega_m$, namely that the halo
MACHO fraction is universal.  According to Fukugita et al. (1998), roughly half
of the baryons are accounted for at redshift zero, so the density of baryonic
MACHOs should not exceed $\Omega_b/2\approx 0.01h^{-2}$, and thus $f\le 1/15$
for $\Omega_m=0.3$ and $h=0.7$.  Of course non-baryonic MACHOs could have
$f=1$.  We will thus take $\Omega_{\rm lens}=0.02$ as the largest plausible
value, though even this value has problems (Graff et al.\ 1999; Fields, Freese
\& Graff 2000).  There are limits on the cosmological abundance of MACHOs of
any kind (Dalcanton et al. 1994), but $\Omega_{\rm lens}<0.1$ is allowed for
stellar--mass lenses.  In the baryonic case, the optical depth to redshift
$z=2$ is as large as 1/75, which is not excessively smaller than the observed
rate.

Though we discussed excising the central region $R<0.2$ of the Einstein ring in
the previous section, to account for afterglow lensing events that happen too
quickly to be detected, doing so has only a small effect on the rate, in fact
this region is only 4\% of the Einstein ring.  This cut will have a larger
effect in the next section, where we will be discussing magnification bias.

The actual optical depth of relevance depends on the magnification one is
interested in.  Assuming $A_0$ is the magnification threshold of interest, the
optical depth is:
\begin{equation}
\tau=\tau_0\int_0^\infty du\,\Theta\left[A_{\rm
max}\left(\sqrt{u}\right)-A_0\right]=\tau_0\beta^2.
\end{equation}
where $A_{\rm max}$ is the peak magnification for the given lens system at a
distance $r=\sqrt{u}$ in Einstein units from the burst.  Here, $\beta$ is the
distance from the lens in Einstein units where the peak magnification is $A_0$.
The symbol $\Theta (x)$ denotes a step function ($=1$ if $x \ge 0$, and zero
otherwise).  Ignoring magnification bias, the optical depth $\tau$ is equal to
the rate of microlensed GRBs.  For \GRB, the peak magnification was
approximately $A_0=2$.

In Table~\ref{tab:tau} we give the optical depth to lensing for several
different magnification thresholds and we also consider three cases for the
fractional width of the emitting ring in the gamma ray burst afterglow.  We
find that the consideration of binary lenses typically increases the optical
depth, but not by a large amount (see Appendix~\ref{appdx:binary}).  The two
main effects that we see are that binaries give a larger increase for narrower
source rings, and a larger increase for larger thresholds.  Note that for this
calculation, we only require that the peak magnification exceed the threshold,
and we do not consider the detailed structure of the lightcurve.

In the remainder of this paper, we will assume that $\Omega_m=0.3$, $A_0=2$ and
$w=0.1$, which gives $\tau=1.4\tau_0$.  We will consider both stellar lenses
with $\Omega_{\rm lens}=0.0035$ and MACHOs with $\Omega_{\rm lens}=f\Omega_m$.

For microlensing of point sources, the probability of an impact parameter less
than $\beta$ in Einstein units, and thus a magnification greater than
$A=(2+\beta^2)/\sqrt{\beta^2(4+\beta^2)}$ is simply given by
\begin{equation}
P=1-e^{-\tau_0\beta^2}\approx\tau_0\beta^2.
\label{Ptau0beta}
\end{equation}
We will use the approximate expression, and thus we need to truncate to ensure
that $P\le1$,
\begin{equation}
P=2\tau_0\left(\frac{A}{\sqrt{A^2-1}}-1\right)
\Theta\left(A-\frac{1+2\tau_0}{\sqrt{1+4\tau_0}}\right).
\label{Ptau0}
\end{equation}
The probability density corresponding to this expression is now
\begin{equation}
\frac{dP}{dA}=\frac{2\tau_0}{\sqrt{(A^2-1)^3}}\,
\Theta\left(A-\frac{1+2\tau_0}{\sqrt{1+4\tau_0}}\right).
\label{PAtau0}
\end{equation}
i.e. $(dP/dA) dA$ gives the probability that the magnification lies in the
range $A \pm dA/2$.  We will briefly discuss the effects of an inhomogeneous
background mass distribution in \S\ref{discussion}.

\section{Single Magnification Bias}
\label{bias}

We now consider the magnification bias (ignoring the complication of binary
lenses, which as we have seen does not significantly affect the lensing rate).
We compute the probability of a detectable microlensing event on the afterglow
of a GRB detected at high energy.  The detectability of the GRB at high energy
will be affected by magnification bias as a point source.  We then simply model
the microlensing event as a detected GRB with a lens between 0.2 and 1.2
Einstein radii distant.  The inner boundary allows a minimum time before peak
magnification (so that a bump in the lightcurve can be observed), and the outer
boundary simply enforces that the magnification is in excess of roughly 2 in
the peak (for $w=0.1$).  The precise values of these parameters do not
significantly affect our results, however, for luminosity functions with slopes
$\gamma<3$.  Implicit in this prescription is the assumption that the
magnification of the the gamma ray flux is the same as the magnification of the
optical flux, when the optical flux is first detected.  This is a good
assumption as typically the afterglow would be detected in the regime where the
emission region is small compared to the Einstein radius, meaning the afterglow
is effectively a point source when initially detected.\footnote{An implicit
assumption in this section is also that the optical luminosity and gamma-ray
luminosity of a burst are well-correlated.  In other words, in order for a
burst to be detected in gamma-ray, {\it and} its afterglow to be observed in
the optical, it is sufficient to consider the luminosity function in one
waveband (in other words, a threshold in one waveband translates directly into
a threshold in another waveband).  We will consider in the next section the
opposite case where the optical and gamma-ray flux are completely uncorrelated.
Note also that, typically, observations of optical afterglows also require
localization via X-ray (using BeppoSAX), making the selection process even more
complicated. However, there are reasons to believe that X-ray and gamma-ray
luminosities are well-correlated (Granot \& Sari 2002).}

We now derive expressions for the number of detected sources, given lensing
effects, following the discussion in Turner, Ostriker \& Gott (1984).  We
assume a simple selection function $S(F)$ which gives the efficiency of
detecting sources with flux $F$, and an absolute luminosity function $N(L)
\equiv dn/dL$, where $F(L,z)= L/(4\pi D_L^2 (z))$, and $N(L)$ is the number of
sources per luminosity range.  In this section and the next, we will consider
bursts at a fixed redshift, thus the apparent luminosity function is trivially
related to the true luminosity function.  Without lensing, the number of
detected sources is then
\begin{equation}
N_0=\int dL S(F[L,z]) N(L).
\label{N0}
\end{equation}
Taking into account lensing, the number of detected sources is
\begin{equation}
N_{\rm lens}=\int dL S(F[L,z])\int
\frac{dA}{A}\,\,\frac{dP}{dA}\,N\left(\frac{L}{A}\right).
\label{Nlens}
\end{equation}
We will consider a simple power--law luminosity function, with a low-- and
high--flux cutoffs,
\begin{equation}
N(L) \propto L^{-\gamma}\Theta(L-\LL)\Theta(\LH-L),
\end{equation}
and a simple choice of selection function\footnote{The realistic selection
function is almost certainly more complex than a simple step function, and
probably never 100\% efficient even far above threshold.}
\begin{equation}
S(F)=\Theta(F-\Ft),
\label{SF}
\end{equation}
where $F(\LH)\equiv\FH\gg\Ft\gg F(\LL)\equiv\FL$.  We define $\fH=\Ft/\FH$ and
$\fL=\FL/\Ft$ (we expect $\fH,\fL\ll1$).  Here we note that if the slope is in
the range $1<\gamma<3$, the cutoffs can be neglected.\footnote{The $\gamma = 1$
limit comes from the convergence of the integrated luminosity function ($\int
dL\,N(L)$) at the low luminosity end, while the $\gamma = 3$ limit arises from
the the convergence of $\int dA \,(dP/dA)\,N(L/A)/A$ in the large $A$ limit,
where $dP/dA \propto A^{-3}$.}  With these simple forms, we find the following
expression for the unlensed source count, which of course is not measurable,
\begin{equation}
N_0 \propto \frac{(4\pi \Ft D_L^2)^{1-\gamma}}{\gamma-1}
\left(1-\fH^{\gamma-1}\right).
\end{equation}

Not explicitly stated in Eq.~\ref{Nlens} is the fact that one is generally
interested in the number of sources with magnification larger than some value,
or in some particular range. We define the quantity
\begin{equation}
B (A > A_0) \equiv N_{\rm lens} (A > A_0) / N_0
\label{BA0}
\end{equation}
which is equal to the number of sources with magnification $A > A_0$,
normalized by $N_0$, the total number of detected sources if lensing were
absent (Eq.~\ref{N0}).  With the power-law luminosity function, this takes
the following form
\begin{equation}
B(A>A_0)=\frac{1}{1-\fH^{\gamma-1}}\int_{A_0}^\infty dA\,\frac{dP}{dA}
\left[\max\left(A^{-1},\fL\right)^{1-\gamma}-\fH^{\gamma-1}\right].
\end{equation}

We can compute $B(A>A_0)$ explicitly for the analytic $P(A)$
(Eq.~\ref{Ptau0}).  First, we will need the integral over the analytic
probability density of an arbitrary power of amplification, given by
\begin{equation}
\int_{A_0}^\infty\frac{dA\;A^{\gamma-1}}{\sqrt{(A^2-1)^3}}=
{\cal F}\left(\frac{3-\gamma}{2};A_0^{-1}\right),
\end{equation}
and we have used the definition
\begin{equation}
{\cal F}(a;x)=
\frac{x^{2a}}{2a}\hypgeo\left(\frac{3}{2},a;a+1;x^2\right)
\end{equation}
with $\hypgeo$ being the standard hypergeometric function.  We notice that this
function has simple poles when $a$ is a non-positive integer, and their
residues are independent of $x$, thus when two such functions are subtracted,
all poles vanish.

We define the quantity $B_0 \equiv B(A > 1)$, which gives us the ratio of the
number of all detected sources taking into account lensing (with all possible
magnification) to the number of all detected sources ignoring lensing.  Note
that by virtue of the approximation made in Eq.~\ref{PAtau0}, we approximate
$B(A > 1)$ by $B(A > [1+2\tau_0]/\sqrt{1+4 \tau_0})$.\footnote{We have
numerically checked that using the exact expression for probability $P = 1 -
e^{-\tau_0 \beta^2}$ instead of $P \sim \tau_0 \beta^2$ (see
Eq.~\ref{Ptau0beta}) results in negligible changes to our results.}  Expanding
$B_0$ in the limit of $\tau_0 \ll 1$, we find
where we have used the following expansion of ${\cal F}$ in the limit of small
$\tau_0$:
\begin{equation}
{\cal F}\left(a;\frac{\sqrt{1+4\tau_0}}{1+2\tau_0}\right)=
\frac{1}{2\tau_0}+1-\frac{\sqrt{\pi}\,\,\Gamma(a)}{\Gamma(a\!-\!1/2)}+
2(a-1)\tau_0+O(\tau_0^2).
\end{equation}

\begin{figure}
\epsfig{width=\textwidth,file=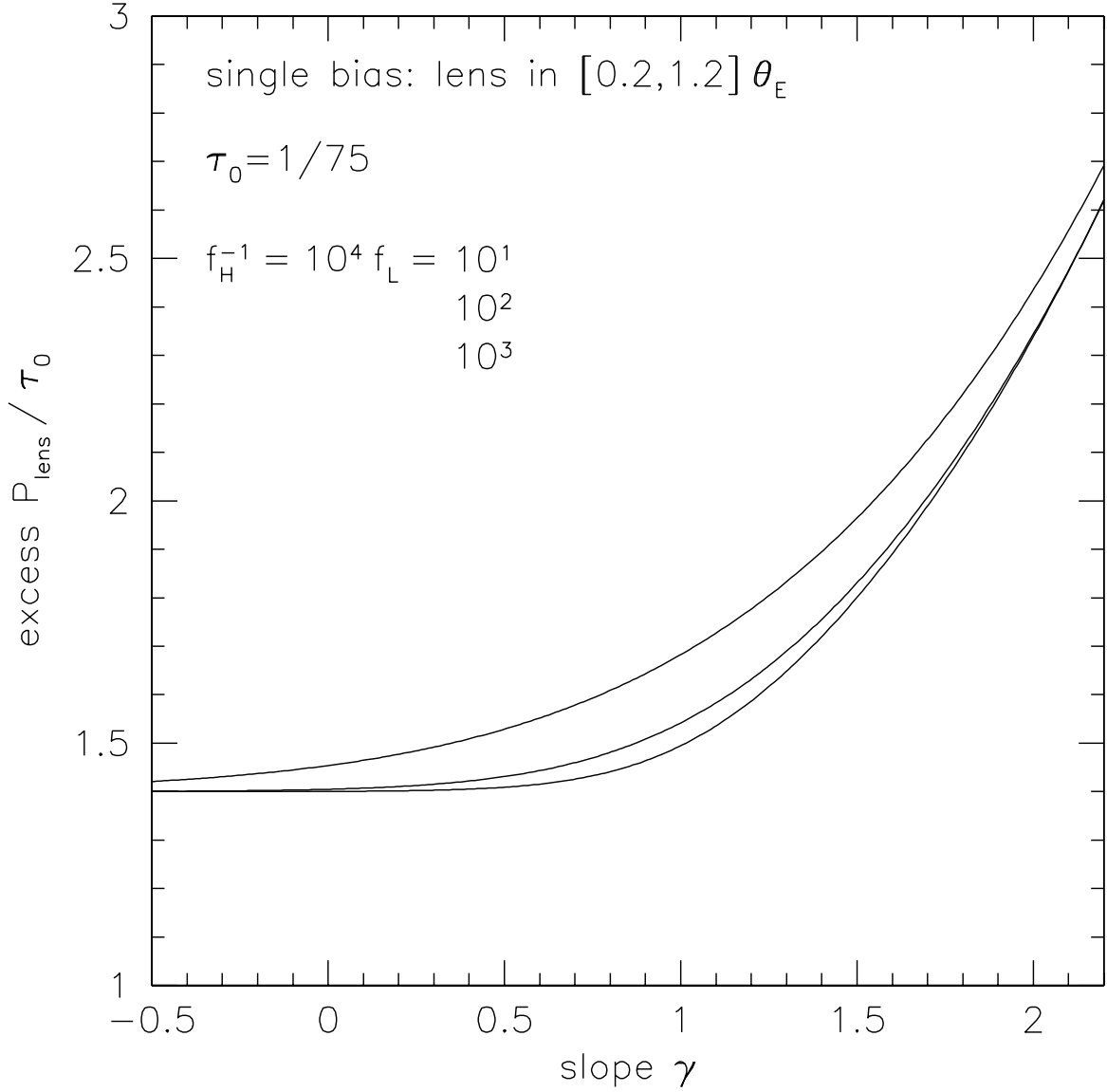}
\caption{Magnification bias as a function of the luminosity function slope.  We
have illustrated three possibilities for the high-- and low--luminosity cutoff.
The value of this cutoff is not very significant when $\gamma$ is not near
three.  The curves are, top to bottom, $\fH=0.1,0.01,0.001$, and we thus see
that the effect of magnification bias is largest when the threshold is not too
far from the maximum luminosity ($\fH$ not too much smaller than unity).}
\label{fig:single}
\end{figure}

We are interested in the number of microlensed sources, which we defined
previously as sources falling in an annulus about a lens, with inner and outer
radii of 0.2 and 1.2 in Einstein units, respectively.
The number of detected microlensed sources is then just the difference of the
bias factors $B$ for minimum magnifications corresponding to the inner and
outer edges of the annulus.  Thus we find that the ratio of the number of
microlensed sources to the number of detected sources, or equivalently the
probability of a given burst being microlensed, is given by
\begin{equation}
P_{\rm lens}=\frac{B(A > A(1.2))-B(A > A(0.2))}{B_0}\sim{\rm several}\;\tau_0,
\label{eq:Plens}
\end{equation}
where the most naive expectation (ignoring magnification bias but taking into
account lensing) is that $P_{\rm lens}= (1.2^2 - 0.2^2) \tau_0 \sim 1.4\tau_0$.
In Fig.~\ref{fig:single} we plot $P_{\rm lens}/\tau_0$ as a function of
$\gamma$ for several values of $\fL$ and $\fH$, with the constraint
$\fL\fH=10^{-4}$, in other words the luminosity function ranges over four
orders of magnitude.\footnote{We assume that $f_{\rm L}$ and $f_{\rm H}$,
defined after Eq.~\ref{SF}, are both small. Their exact size, as can be seen
from Fig.~\ref{fig:single}, does not matter a great deal.  We choose the range
$\fL\fH$ to span four orders of magnitude as Schaefer et al. (2001) give the
luminosity function over this range.  If $\fL\fH$ were larger, our results are
not much affected.}  We have used the magnification distribution derived from
the distribution in $\kappa$, though the results do not differ much from those
of the analytic $dP/dA$.  We see that the magnification bias is only
significant when the slope $\gamma$ becomes significantly larger than one.
Note that we have fixed $\tau_0=1/75$ in Fig.~\ref{fig:single}, but the curves
depend on the value of $\tau_0$ only weakly.  In other words, the lensing
probability is nearly linear in $\tau_0$.

Using the value of $\gamma = 1.7$ for the faint-end luminosity function from
Schaefer et al. (2001), we find the magnification bias as a function of the
ratios $\fL$ and $\fH$,
\begin{equation}
\frac{P_{\rm lens}}{\tau_0}=\frac{2.016-1.400\fH^{0.7}}
{1-\fH^{0.7}+\tau_0(1.211-0.538\fL^{1.3})}.
\label{eq:plens}
\end{equation}
One can see that if $\fL$ and $\fH$ can be ignored, $P_{\rm lens}/\tau_0 \sim
2$.  Even with this slope significantly less than three, the magnification bias
causes an enhancement of the probability of a factor of roughly fifty percent.
We should emphasize that magnification bias offers a smaller enhancement if
$\gamma$ were smaller.

\section{Double Magnification Bias}
\label{doublebias}

In the previous section we effectively assumed that the gamma-ray / X-ray and
optical luminosities were perfectly correlated.  We now take the other extreme
position, assuming that the gamma-ray / X-ray and optical luminosities are
completely uncorrelated, though, for simplicity, both drawn from luminosity
functions with the same shape but with different thresholds (see discussion in
\S \ref{physics}).  It has been shown that this situation can give rise to
significantly larger magnification bias (Borgeest, van Linde \& Refsdal 1991;
Wyithe, Winn \& Rusin 2002).  As before, the unlensed number counts are given
by
\begin{equation}
N_0=\int dL_V\,S_V (F [L_V, z]) N(L_V)\int dL_X\,S_X (F [L_X, z]) N(L_X).
\end{equation}
where the subscripts $V$ and $X$ stand for the visible and X-ray wavebands
relevant for selection of sources. Note that here $dL_V dL_X N(L_V) N(L_X)$
gives the number of sources with the luminosities in the two wavebands falling
into the respective ranges.  Taking into account lensing, the number of
detected sources is
\begin{equation}
N_{\rm lens}=\int dL_V\,S_V (F[L_V, z])\int dL_X\,S_X (F[L_X, z])
\int \frac{dA}{A^2}\,\,\frac{dP}{dA}\,
N\left(\frac{L_V}{A}\right)N\left(\frac{L_X}{A}\right).
\end{equation}
We will now define some similar variables to the previous section.  We take the
luminosity functions in the two bands to be identical, with the same slope and
range between lower and upper cutoffs, though the detection thresholds
differ.\footnote{ Theoretically, the cutoffs for one waveband: $L_{\rm L}^V$
and $L_{\rm H}^V$, and the cutoffs for another: $L_{\rm L}^G$ and $L_{\rm
H}^G$, can be completely different. What we assume here is that the range
$L_{\rm H}^V / L_{\rm L}^V$ equals $L_{\rm H}^G/ L_{\rm L}^G$. Our expressions
here implicitly assume a rescaling has been done so that $L_{\rm L}^V = L_{\rm
L}^G$ and $L_{\rm H}^V = L_{\rm H}^G$. The rescaling factors are omitted from
our expressions, because they get canceled out in the end as far as the
probability of lensing is concerned.}  We define $\fL$ and $\fH$ as before,
based on the {\em smaller} detection threshold (relative to the lower
luminosity function cutoff), and we define $\eta\le1$ as the ratio of the
smaller to larger threshold.  The unlensed source counts are given by
\begin{equation}
N_0 = \frac{(4\pi \Ft D_L^2)^{2-2\gamma}}{(\gamma-1)^2}
\left(1-\fH^{\gamma-1}\right)\left(\eta^{\gamma-1}-\fH^{\gamma-1}\right),
\end{equation}
where $\Ft$ is now the lower of the two thresholds.  Similarly to the previous
section, we compute the excess source counts above a given magnification $A_0$
due to the double magnification bias, $B(A > A_0) \equiv N_{\rm lens} (A >
A_0)/N_0$ (Eq.~\ref{BA0}):
\begin{eqnarray}
B(A>A_0)&=&\left(1-\fH^{\gamma-1}\right)^{-1}
\left(\eta^{\gamma-1}-\fH^{\gamma-1}\right)^{-1}\times\\
&&\int_{A_0}^\infty dA\,\frac{dP}{dA}
\left[\max\left(A^{-1},\fL\right)^{1-\gamma}-\fH^{\gamma-1}\right]
\left[\max\left(\eta^{-1}A^{-1},\fL\right)^{1-\gamma}-\fH^{\gamma-1}\right].
\nonumber
\end{eqnarray}

\begin{figure}
\epsfig{width=\textwidth,file=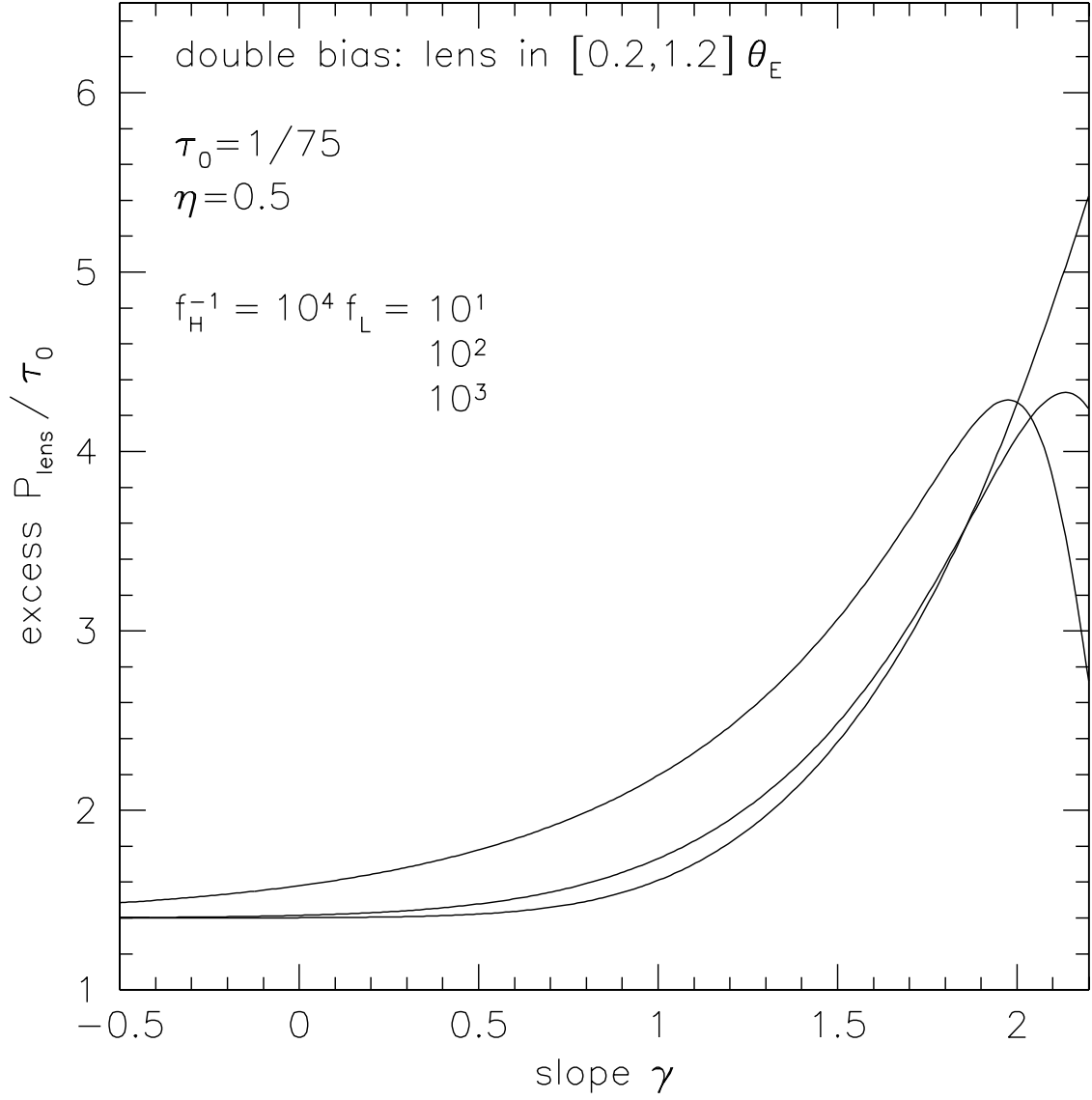}
\caption{Double magnification bias as a function of the luminosity function
slope.  We have used the same three possibilities for the high-- and
low--luminosity cutoff as in Fig.~\ref{fig:single}.  The value of this cutoff
is not very significant unless $\gamma$ is larger than two.  The curves match
those in Fig.~\ref{fig:single}.}
\label{fig:double}
\end{figure}

We now repeat the analysis of the previous section for the double magnification
bias, with the same range of allowed magnifications, again using
Eq.~\ref{eq:plens}.  We plot the ratio as before in Fig.~\ref{fig:double},
where we again fix $\tau_0=1/75$.  We will take $\eta=0.5$, and the results are
insensitive to the exact value.  We see that the magnification bias is
insignificant for the case where $\gamma < 1$, but it can be as large as three
or four in the case where $\gamma=1.7$ (motivated by the results of Schaefer et
al. 2001; see \S \ref{physics} for more discussions).  Note that, however, if
$\gamma$ gets close to $2$ or larger, the enhancement in lensing probability
due to double magnification bias can become very large, limited only by the
cut-offs in the luminosity functions -- i.e. the result will be quite cut-off
dependent.  In fact in some cases a downturn is expected, as the very high
magnification regime yields microlensing events too short to be detected.
Lastly, if the two luminosity function slopes are different, the important
quantity is their average, so if $(\gamma_1+\gamma_2)/2\rightarrow 2$, then the
double magnification bias becomes large.

\section{Afterglow Sample}
\label{sample}

\begin{figure}
\epsfig{width=\textwidth,file=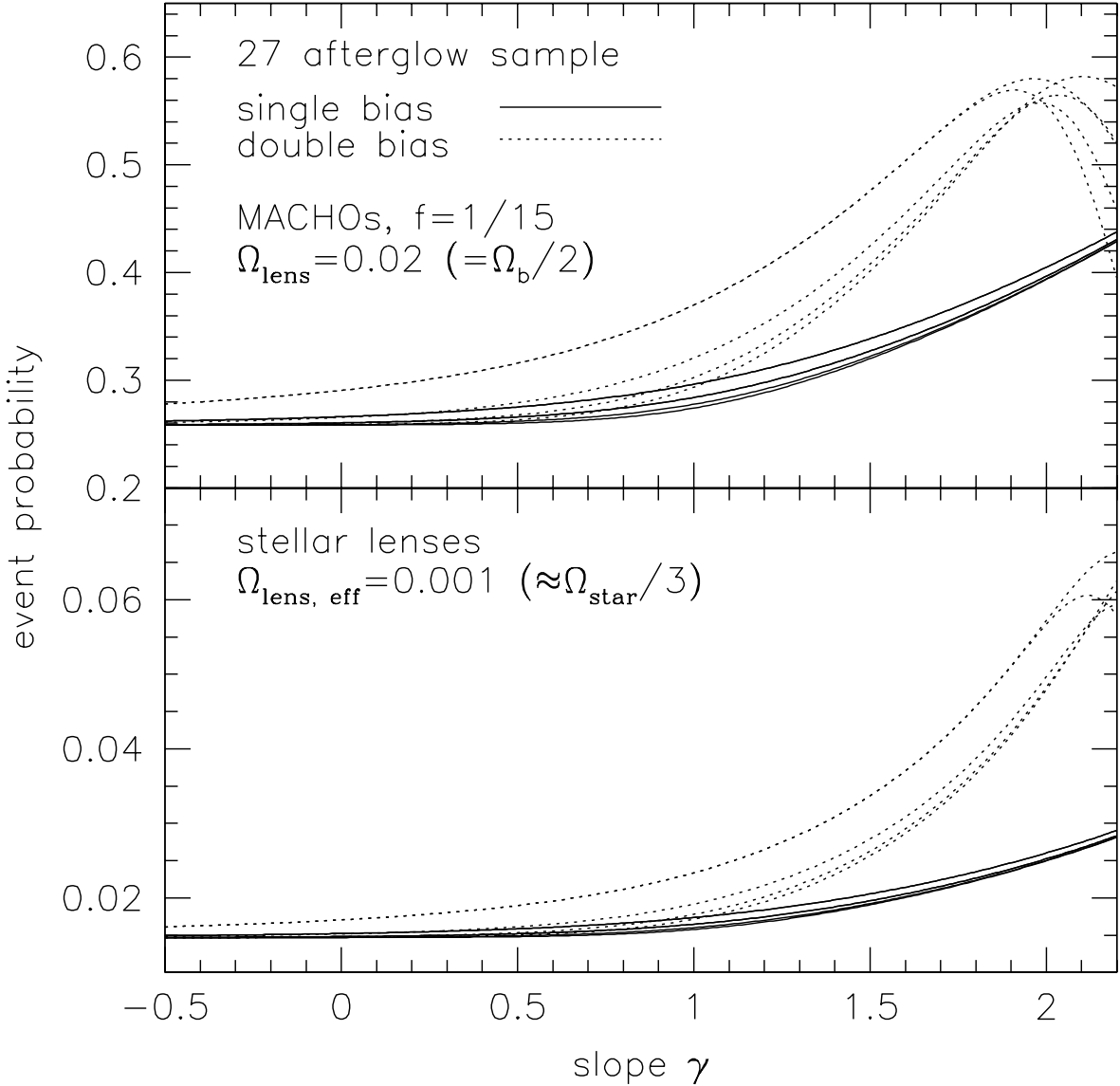}
\caption{Probability of at least one microlensing event in the sample.  27
afterglows with well--determined redshifts are considered, adjusting the
optical depth and thresholds according to cosmology only.  Several curves are
shown for both the single and double magnification bias cases.  In the top
panel, we illustrate MACHO lensing where $\Omega_{\rm lens}=0.02$, namely half
of the baryons in the universe.  In the bottom panel, we illustrate stellar
lensing, accounting for the fact that stars are less effective per mass as they
are strongly clustered (Wyithe \& Turner 2002a; Koopmans \& Wambsganss 2001).
This is a somewhat crude estimate, and accordingly we have used the analytic
$dP/dA$.  As in Figs.~\ref{fig:single}-\ref{fig:double}, the higher probability
curves correspond to the largest $\fH$ (though still small compared with
unity).  Changing the range of luminosity function ($\fL\fH=10^{-4},10^{-5}$)
has only a small effect in that the excess probability function turns over at
slightly smaller slopes for the $\fL\fH=10^{-5}$ case.}
\label{fig:sample}
\end{figure}

The results of the previous two sections are valid for bursts at a single
redshift only.  To be more precise, we will consider a population of afterglows
from a range of redshifts.  We choose 27 detected afterglows as our sample,
taken from Bloom et al. (2003) and listed in Table~\ref{tab:GRB}.  While we
caution that the sample is a heterogeneous one, with a probably quite complex
selection process, we will continue to model only the part of the selection
that has to do with the observed flux (in gamma-ray as well as in optical) --
that our results are not sensitive to the precise thresholds chosen (unless
$\gamma$ is very close to $3$ or larger for single magnification bias, or $2$
or larger for double magnification bias) will be regarded as a partial
justification for our model.  We calculate the microlensing optical depth as a
function of redshift, as before assuming the flat $\Omega_m=0.3$ cosmology, and
assuming that the shape of the luminosity function does not evolve.  For each
redshift, we compute the probability of a detectable microlensed GRB.  For the
full sample, we then simply compute the probability that at least one observed
afterglow was microlensed.  In Fig.~\ref{fig:sample} we show results for
various values of the thresholds, and for both $\fL\fH=10^{-4}$ and
$\fL\fH=10^{-5}$.  In principle we could use the flux information for each
burst to construct a probability, but as we show in Appendix~\ref{appdx:flux},
this would make little difference.

We have thus far assumed that lenses are uniformly distributed (i.e.\ not
clustered), an approximation that can fail badly.  For lenses distributed in
isothermal galactic halos, as MACHOs would be Wyithe \& Turner (2002a) found
that the typical cosmological microlensing event occurs at small optical depth,
and thus the isolated lens approximation is justified.  In the top panel of
Fig.~\ref{fig:sample} we illustrate the probability to see at least one event
in the sample assuming $f=1/15$ (MACHOs making up half of $\Omega_b$).  This
probability is quite high: about 25\% with no magnification bias and
approaching 35\% and 50\% for single and double magnification bias with
$\gamma=1.7$, respectively.  These probabilities make it likely that another
microlensed GRBs could be observed if the sample were to be doubled.

In contrast, Wyithe \& Turner (2002a) found that microlensing by stars
typically occurs at optical depths of order unity, and the isolated lens
approximation is not appropriate.  Koopmans \& Wambsganss (2001) have addressed
this problem by simulating lens systems at optical depths of order unity.  They
find a microlensing probability proportional to optical depth for small optical
depths, but turning over at optical depths around 1/10.  Choosing a typical
value of $\tau=1/4$ according to Wyithe \& Turner (2002a), the effectiveness of
lenses is of order only one third the naive value.  In discussing stellar
lenses we will thus take an effective lens density $\Omega_{\rm
lens,eff}\approx 10^{-3}$.  In the bottom panel of Fig.~\ref{fig:sample} we
illustrate the probability of observing at least one GRB lensing event due to a
star in the sample.  Without magnification bias, this is about 1.5\%, rising to
2\% and 3.5\% for single and double magnification bias respectively.  It thus
seems unlikely that a stellar lens could be responsible.

\section{Discussion and Conclusions}
\label{discussion}

Our estimate of the microlensing rate for gamma ray burst afterglows is in
significant disagreement with previous literature, which we outline here.  If
$f=1$, then the optical depth to $z=2$ is $\tau_0\approx0.2$, in significant
disagreement with Koopmans and Wambsganss (2001).  Computing the mean value of
$\kappa$ from their Eq.~4 (this is identified with our $\tau_0$), we find
$\overline{\kappa}\approx0.04$, a factor of five smaller.  They have cut their
isothermal halos at radii that are too small, and have thus neglected 80\% of
their mass.  They have done this to enforce that the halos do not overlap along
the line of sight, but the fact is that they do, and significantly.  They have
thus underestimated the lensing probability by a factor of five.  Let us
explain this in more detail.

In \S\ref{microlensing} we assumed an optical depth for microlensing consistent
with a uniform distribution of lenses.  This is clearly not the case, as mass
in clustered in halos.  We assume that all of the matter is in singular
isothermal sphere halos with cutoffs ($\rho(r)=\sigma^2/(2\pi
Gr^2)\Theta(r_0-r)$).  These will obey a luminosity--velocity dispersion
relation, have luminosities taken from a Press-Schechter distribution, and have
relation between cutoff radii and velocity dispersion,
\begin{eqnarray}
L^\star\,\frac{dn}{dL}&=&n^\star\left(\frac{L}{L^\star}\right)^\alpha\,
e^{-L/L^\star},\\
\frac{L}{L^\star}&=&\left(\frac{\sigma}{\sigma^\star}\right)^\gamma,\\
\frac{r_0}{r_0^\star}&=&\left(\frac{\sigma}{\sigma^\star}\right)^\beta.
\end{eqnarray}
Following Koopmans \& Wambsganss (2001), we will take $\alpha=-1$, $\gamma=4$,
as the fiducial case, with $n^\star=0.0061\,h^3$ Mpc$^{-3}$, $\sigma^\star=225$
km s$^{-1}$.  To prescribe the cutoff radii, we take $\beta=1$, but the results
are insensitive to the precise value of beta as long as it is not close to
zero.  We will compute $r_0^\star$ by normalizing the total mass density to
account for $\Omega_m=0.3$, namely all mass lies in halos.  We find
\begin{equation}
r_0^\star=\left(\frac{3}{16\pi}\right)
\frac{\Omega_m H_0^2}{n^\star\sigma^{\star2}}\;
\Gamma\left(1+\alpha+\frac{2+\beta}{\gamma}\right)^{-1}=0.47
\;h^{-1}\;{\rm Mpc}.
\end{equation}

We now can easily compute the geometric cross section of halos, and the mean
number of halos intersected on a line of sight to a particular redshift,
\begin{eqnarray}
\tau_{\rm GEOM}&=&\left(\frac{2\pi}{3}\right)
\left(\frac{n^\star r_0^{\star2}c}{\Omega_mH_0}\right)
\Gamma\left(1+\alpha+2\beta/\gamma\right)\;
\left(\sqrt{\Omega_m(1+z)^3+\Omega_\Lambda}-1\right),\\
&=&50.7\left(\sqrt{\Omega_m(1+z)^3+\Omega_\Lambda}-1\right).
\end{eqnarray}
Even to moderate redshift, a significant number of halos are intersected.  For
\GRB\ at $z=2.034$, $\tau_{\rm GEOM}\approx 100$, and the approximation that
halos do not overlap is therefore not accurate.  This is the origin of our
disagreement.

We have shown that a large number of halos are typically intersected.  However,
there will be variations in the optical depth (or equivalently, convergence)
between lines of sight.  We have done a simple test, simulating the
distribution of convergences and modifying the magnification distribution
$dP/dA$ accordingly.  The magnification bias we find does not vary by more than
15\% in this case.

We have shown that the consideration of magnification bias (both single and
double) is important, in that the expected probability that a given observed
gamma ray burst is microlensed can be roughly three times the naive optical
depth estimate.  MACHOs making up half of the baryonic matter have a
significant probability to cause such an event, but we find that stellar lenses
alone have a very small probability (less than 3\% for any event in the
afterglow sample) to do so.  We find that binaries can affect the gamma ray
burst microlensing rate at the 25\% level, which though surprisingly large, is
not very significant.

A single observed event with a low probability is not in itself problematic.
However, should another microlensed gamma ray burst be observed in another
thirty afterglows, we believe it would be a significant indication that there
are significantly more lenses in the universe than can be inferred from stars.
In that case, the effects we have calculated could be quite important in
determining the exact density of lenses.

\acknowledgments

We thank Jules Halpern for numerous enlightening discussions, and Andrei
Gruzinov, Leon Koopmans, Pawan Kumar, Dani Maoz, Re'em Sari and Eli Waxman for
useful conversations.  EAB acknowledges support from the Columbia University
Academic Quality Fund.  LH acknowledges support by an Outstanding Junior
Investigator Award from the DOE, grant AST-0098437 from the NSF and grant
NAG5-10842 from NASA.  This research was supported in part by the NSF under
grant PHY99-07949 at the Kavli Institute for Theoretical Physics at the
University of California, Santa Barbara.

\appendix

\section{Binary Lenses}
\label{appdx:binary}

We discuss the two point mass gravitational lens, a model that has been studied
extensively (Schneider \& Wei\ss\ 1986; Erdl \& Schneider 1993).  We quote some
relevant results.  We take the masses of the lenses to be $M_1$ and $M_2$, with
$M=M_1+M_2$, and we define dimensionless masses $\nu_{1,2}=M_{1,2}/M$ such that
$\nu_1+\nu_2=1$.  Furthermore, we define $q=M_2/M_1$, with $0\le q\le 1$
without loss of generality.  Note that taking $q=0$ yields the single point
mass lens.

The excess optical depth due to binaries is given by
\begin{equation}
\frac{\Delta\tau}{\tau}=\frac{\tau_{2*}-\tau_{1*}}{\tau_{1*}}=
\frac{\beta_{2*}^2}{\beta_{1*}^2}-1.
\end{equation}
For the binary systems, we take a distribution flat in $q\equiv M_2/M_1$, and
also flat in $\log d$, where $d$ is the projected separation of the lenses on
the sky, for the ensemble average (Baltz \& Gondolo 2001).

We adopt a complex parameterization (Witt 1990) of the lens system. We
introduce complex angular coordinates on the plane of the sky, $z = (\theta_x +
i\theta_y)/\theta_E$.  Given two lenses at angular positions $z_1$ and $z_2$, a
source at $\zeta$ will have images at the solutions $z$ of the lens equation
\begin{equation}
\label{eq:zeta}
\zeta=z-\overline{\alpha(z)}=z-\frac{\nu_1}{\overline{z}-\overline{z}_1}-
\frac{\nu_2}{\overline{z}-\overline{z}_2}.
\end{equation}
In this formalism, the Jacobian of $\zeta(z)$ in Eq.~\ref{eq:zeta} is
\begin{equation}
J(z)=1- \left| \frac{\partial\alpha}{\partial z} \right|^2 .
\end{equation}
The magnification of an image at $z$ of a point source at $\zeta$ is given by
$1/|J(z)|$, with the sign of $J(z)$ giving the parity $p$ of the image. The
total magnification is the sum of the individual magnifications of the images.

For a source of uniform surface brightness, the magnification is just the image
area divided by the source area (Gould \& Gaucherel 1997),
\begin{equation}
\mu=
\int_0^{2\pi}d\phi\;{\rm Im}\left[\sum_i p_i\overline{z}_i\frac{dz_i}{d\phi}
\right]\Bigg{/}
\int_0^{2\pi}d\phi\;{\rm Im}\left[\overline{\zeta}\frac{d\zeta}{d\phi}
\right].
\end{equation}
We never need to use numerical differencing as long as we can analytically
differentiate the expression for the source boundary $\zeta(\phi)$.  As we are
concerned with annular sources, the source boundaries will always have the form
$\zeta=\zeta_0+re^{i\phi}$, which can be easily differentiated.
Differentiating the lens equation, we find
\begin{equation}
\frac{d\zeta}{d\phi}=\frac{dz}{d\phi}-\overline{\frac{\partial\alpha}
{\partial z}}\overline{\frac{dz}{d\phi}},
\end{equation}
Solving for $dz/d\phi$, we find the following result, depending only on the
numerical solution of the lens equation, and not on any numerical derivatives
\begin{equation}
\frac{dz}{d\phi}=\frac{1}{J(z)}
\left(\frac{d\zeta}{d\phi}+\overline{\frac{\partial\alpha}{\partial z}}
\overline{\frac{d\zeta}{d\phi}}\right).
\label{eq:dzdphi}
\end{equation}
We can take care of the required parities trivially.  Since we effectively need
$p_i\,dz_i/d\phi$, we simply take the absolute value of the Jacobian factor
$J(z)$ in Eq.~\ref{eq:dzdphi}, since $p_i={\rm sign}\;{J(z_i)}$.

\section{Flux Information}
\label{appdx:flux}

Given that we have the gamma ray fluxes for this sample of bursts, we could
consider using that information in the determination of the probability.  In
particular, we could write the probability for an observed burst with apparent
flux $F_i$ and redshift $z_i$,
\begin{equation}
P_{\rm lens}(F_i,z_i)=\frac{\int_{A_{1.2}}^{A_{0.2}}dA\,(dP/dA)\,
N(L(F_i,z_i)/A)/A} {\int_1^\infty dA\,(dP/dA)\,N(L(F_i,z_i)/A)/A}.
\end{equation}
This compares with Eq.~\ref{eq:Plens} for the single magnification bias,
which can be written in a similar form
\begin{equation}
P_{\rm lens}(z_i)=
\frac{\int_{A_{1.2}}^{A_{0.2}}dA\,(dP/dA)/A
\int_\Ft^\infty dF\,N(L(F,z_i)/A)}
{\int_1^\infty dA\,(dP/dA)/A
\int_\Ft^\infty dF\,N(L(F,z_i)/A)}.
\end{equation}
For a power law luminosity function with no cutoffs, as is approximately the
case when $1<\gamma<3$, these two probabilities are identical burst by burst.
Summing over the sample reduces the small discrepancy between them when the
luminosity function is not a pure power law.

\begin{deluxetable}{ccccccccccccc}
\tablecaption{Optical Depth for Binary Lensing of GRBs\label{tab:tau}}
\tablehead{
threshold & & \multicolumn{3}{c}{$w=0.20$} & & \multicolumn{3}{c}{$w=0.10$} & &
\multicolumn{3}{c}{$w=0.05$}\\
magnification & & $\beta_{1*}^2$ & $\beta_{2*}^2$ & $\Delta\tau/\tau$ & &
$\beta_{1*}^2$ & $\beta_{2*}^2$ & $\Delta\tau/\tau$ & &
$\beta_{1*}^2$ & $\beta_{2*}^2$ & $\Delta\tau/\tau$
}
\startdata
5.0 & &
0.135 & 0.142 & 4.93\% & & 0.145 & 0.179 & 23.6\% & & 0.185 & 0.259 & 40.0\%\\
3.0 & &
0.341 & 0.359 & 5.22\% & & 0.459 & 0.525 & 14.4\% & & 0.599 & 0.763 & 27.3\%\\
2.0 & &
1.007 & 1.134 & 1.77\% & & 1.399 & 1.503 & 7.47\% & & 1.849 & 2.121 & 14.7\%\\
1.5 & &
2.785 & 2.793 & 0.30\%\tablenotemark{a} & & 4.022 & 4.111 & 2.21\% & &
5.534 & 5.901 & 6.63\%
\enddata

\tablenotetext{a}{The Monte Carlo errors in the computed values of
$\Delta\tau/\tau$ are as large as a few tenths of a percent, thus this result
is marginally consistent with no enhancement due to binary lenses.}

\tablecomments{Optical depth is given by $\tau=\tau_0\beta^2_{n*}$.  The values
of $\beta_{2*}$ are calculated assuming that binary systems with separations
differing from unity by more than half of an order of magnitude act as single
lenses.  We thus slightly underestimate the effect of binary lenses.}
\end{deluxetable}

\begin{deluxetable}{lccccc}
\tablecaption{Sample of GRB Afterglows\label{tab:GRB}}
\tablehead{
GRB & redshift & $\chi$ & $(H_0D_L/c)^2$ & Approximate Fluence  &
Scaled Fluence\\
 & & & &($10^{-6}$ erg cm$^{-2}$) &($10^{-6}$ erg cm$^{-2}$) }
\startdata
970228 & 0.695 & 0.5817 & 0.972 & 11.0 & 10.69\\
970508 & 0.835 & 0.6730 & 1.525 & 3.17 & 4.834\\
970828 & 0.958 & 0.7473 & 2.141 & 96.0 & 205.5\\
971214 & 3.418 & 1.5712 & 48.19 & 9.44 & 454.9\\
980613 & 1.096 & 0.8245 & 2.987 & 1.71 & 5.108\\
980703 & 0.966 & 0.7519 & 2.185 & 22.6 & 49.38\\
990123 & 1.600 & 1.0609 & 7.608 & 268. & 2040.\\
990506 & 1.300 & 0.9280 & 4.556 & 194. & 883.9\\
990510 & 1.619 & 1.0687 & 7.834 & 22.6 & 177.0\\
990705 & 0.840 & 0.6762 & 1.548 & 93.0 & 144.0\\
990712 & 0.433 & 0.3888 & 0.310 & 6.50 & 2.015\\
991208 & 0.706 & 0.5891 & 1.010 & 100. & 101.0\\
991216 & 1.020 & 0.7827 & 2.500 & 194. & 485.0\\
000131 & 4.500 & 1.7496 & 92.60 & 41.8 & 3871.\\
000210 & 0.846 & 0.6801 & 1.577 & 61.0 & 96.20\\
000301C & 2.034 & 1.2208 & 13.72 & 4.10 & 56.25\\
000418 & 1.119 & 0.8367 & 3.143 & 20.0 & 62.86\\
000911 & 1.059 & 0.8041 & 2.740 & 230. & 630.2\\
000926 & 2.037 & 1.2218 & 13.77 & 6.20 & 85.37\\
010222 & 1.477 & 1.0089 & 6.244 & 120. & 749.3\\
010921 & 0.451 & 0.4029 & 0.342 & 15.4 & 5.267\\
011121 & 0.362 & 0.3312 & 0.203 & 24.0 & 4.872\\
011211 & 2.140 & 1.2552 & 15.53 & 5.00 & 77.65\\
020405 & 0.690 & 0.5782 & 0.955 & 38.0 & 36.29\\
020813 & 1.254 & 0.9057 & 4.167 & 38.0 & 158.3\\
021004 & 2.332 & 1.3135 & 19.15 & 3.20 & 61.28\\
021211 & 1.006 & 0.7748 & 2.416 & 1.00 & 2.416
\enddata

\tablecomments{Redshifts, coordinate distance, luminosity distance, and
observed and scaled fluence are given.  The redshifts and fluences are taken
from Bloom et al. (2003).}
\end{deluxetable}


\begin{references}

\reference{baltz} Baltz, E. A. \& Gondolo, P. 2001, \apj, 559, 41

\reference{bloom} Bloom, J. S., Frail, D. A. \& Kulkarni, S. R. 2003, \apj,
594, 674

\reference{double} Borgeest, U., van Linde, J. \& Refsdal, S. 1991, \aap, 251,
L35

\reference{dalcanton} Dalcanton, J. J., Canizares, C. R., Granados, A.,
Steidel, C. C. \& Stocke, J. T. 1994, \apj, 424, 550

\reference{Erdl} Erdl, H. \& Schneider, P. 1993, \aap, 268, 453


\reference{freese2} Fields, B.~D., Freese, K. \& Graff, D.~S.\ 2000, \apj, 534,
265

\reference{Frail} Frail, D. A. et al. 2001, \apj, 562, L55

\reference{FW} Freedman, D. L. \& Waxman, E. 2001, \apj, 547, 922

\reference{budget} Fukugita, M., Hogan, C. J. \& Peebles, P. J. E. 1998, \apj,
503, 518

\reference{GLS} Garnavich, P. M., Loeb, A. \& Stanek, K. Z. 2000, \apj, 544,
L11

\reference{freese1} Graff, D.~S., Freese, K., Walker, T.~P. \& Pinsonneault,
M.~H.\ 1999, \apjl, 523, L77

\reference{granot} Granot, J.~\& Loeb, A. 2001, \apj, 551, L63

\reference{ring} Granot, J., Piran, T. \& Sari, R. 1999, \apj, 527, 236

\reference{corr} Granot, J. \& Sari, R. 2002, \apj, 568, 820

\reference{gaudi1} Gaudi, B.~S., Granot, J. \& Loeb, A. 2001, \apj, 561, 178

\reference{gaudi2} Gaudi, B.~S. \& Loeb, A. 2001, \apj, 558, 643

\reference{ioka} Ioka, K. \& Nakamura, T. 2001, \apj, 561, 703

\reference{koopmans} Koopmans, L.~V.~E. \& Wambsganss, J. 2001, \mnras, 325,
1317

\reference{KTH} Krumholz, M., Thorsett, S. E. \& Harrison, F. A. 1998, \apj,
506, L81

\reference{kulkarni} Kulkarni, S. R., et al. 1998, \nat, 393, 35

\reference{kumar} Kumar, P.\ 2000, \apj, 538, L125

\reference{lp} Loeb, A. \& Perna, R. 1998, \apj, 495, 597

\reference{madau1} Madau, P., Ferguson, H.~C., Dickinson, M.~E., Giavalisco,
M., Steidel, C.~C. \& Fruchter, A. 1996, \mnras, 283, 1388

\reference{madau2} Madau, P., Pozzetti, L., \& Dickinson, M. 1998, \apj, 498,
106

\reference{MaoLoeb} Mao, S. \& Loeb, A. 2001, \apj, 547, L97

\reference{metzger} Metzger, M. R., Djorgovski, S. G., Kulkarni, S. R.,
Steidel, C. C., Adelberger, K. L., Frail, D. A., Costa, E. \& Fronterra,
F. 1997, \nat, 387, 879

\reference{panaitescu} Panaitescu, A. 2001, \apj, 556, 1002

\reference{pressgunn} Press, W.~H.~\& Gunn, J.~E.\ 1973, \apj, 185, 397

\reference{RHL} Rutledge, R.~E., Hui, L., \& Lewin, W.~H.~G. 1995, \mnras, 276,
753

\reference{sari} Sari, R., Piran, T., \& Narayan, R.\ 1998, \apj, 497, L17

\reference{schaefer1} Schaefer, B. E., Deng, M. \& Band, D. L. 2001, \apj, 563,
L123

\reference{Schneider} Schneider, P. \& Wei\ss, A. 1986, \aap, 164, 237

\reference{bias1} Turner, E. L., Ostriker, J. P. \& Gott, J. R. 1984, \apj,
284, 1

\reference{waxmanA} Waxman, E. 1997a, \apj, 489, L33

\reference{waxmanB} Waxman, E. 1997b, \apj, 491, L19

\reference{wijers} Wijers, R.~A.~M.~J., Bloom, J.~S., Bagla, J.~S. \&
Natarajan, P.\ 1998, \mnras, 294, L13

\reference{Witt} Witt, H. J. 1990, \aap, 236, 311

\reference{wyithe} Wyithe, J. S. B. \& Turner, E. L. 2002a, \apj, 567, 18

\reference{wyithe1} Wyithe, J. S. B. \& Turner, E. L. 2002b, \apj, 575, 650

\reference{wyithe2} Wyithe, J. S. B., Winn, J. N. \& Rusin, D. 2003, \apj, 583,
58


\end{references}
\end{document}